\documentclass[11pt,preprint]{aastex}
\usepackage{hyperref}

\newcommand{\kms}{\ensuremath{{\rm km~s}^{-1}}}

\newcommand{\fluxw}{W cm$^{-2}$ }

\shorttitle{MIR Spectral Properties of PSQs}
\shortauthors{Wei et al.}

\begin{document}

\title{Mid-infrared Spectral Properties of Post-Starburst Quasars}

\author{Peng Wei\altaffilmark{1},
 Zhaohui Shang\altaffilmark{1,2},
 Michael S.  Brotherton\altaffilmark{2},
 Sabrina L. Cales\altaffilmark{3},
 Dean C.  Hines\altaffilmark{4},
 Daniel A.  Dale\altaffilmark{2},
 Rajib Ganguly\altaffilmark{5},
 Gabriela Canalizo\altaffilmark{6,7}
}

\altaffiltext{1}{Astrophysics Center, Tianjin Normal University, Tianjin
300387, China; zshang@gmail.com}

\altaffiltext{2}{Department of Physics and Astronomy, University of Wyoming,
Laramie, WY 82071, USA}


\altaffiltext{3}{Departamento de Astronom{\'{\i}}a, Universidad de
Concepci{\'{o}}n, Casilla 160-C, Concepci{\'{o}}n, Chile}


\altaffiltext{4}{Space Telescope Science Institute,
3700 San Martin Drive, Baltimore, MD 21218, USA}

\altaffiltext{5}{Department of Computer Science, Engineering, \& Physics,
University of Michigan-Flint, Flint, MI 48502, USA}

\altaffiltext{6}{Department of Physics and Astronomy, University of California,
Riverside, CA 92521, USA}

\altaffiltext{7}{Institute of Geophysics and Planetary Physics, University of
California, Riverside, CA 92521, USA}

\begin{abstract}

We present {\it{Spitzer}} InfraRed Spectrograph (IRS) low-resolution
spectra of 16 spectroscopically selected post-starburst quasars (PSQs)
at z $\sim$\ 0.3.  The optical spectra of these broad-lined active
galactic nuclei (AGNs) simultaneously show spectral signatures of
massive intermediate-aged stellar populations making them good
candidates for studying the connections between AGNs and their hosts.
The resulting spectra show relatively strong polycyclic aromatic
hydrocarbon (PAH) emission features at 6.2 and 11.3\,\micron\
and a very weak silicate feature, indicative of ongoing star formation
and low dust obscuration levels for the AGNs.  
We find that the
mid-infrared composite spectrum of PSQs has spectral properties
between ULIRGs and QSOs, suggesting that PSQs are hybrid 
AGN and starburst systems as also seen in their optical spectra.  
We also find that PSQs in early-type host
galaxies tend to have relatively strong AGN activities, while those in
spiral hosts have stronger PAH emission, indicating more star
formation.  


\end{abstract}


\section{INTRODUCTION\label{sec:introduction}}

Active galactic nuclei (AGNs) are powered by accretion of mass onto
the super-massive black holes (SMBHs), which have also been found to
exist at the centers of essentially all massive non-active galaxies
\citep[e.g., ][]{Kormendy95}.  This leads to the general belief that
AGNs play an important role in the formation and evolution of galaxies
and several tight correlations have been found between the mass
of SMBHs and various properties of host galaxies, such as the velocity
dispersion, \citep{Ferrarese00, Gebhardt00, Tremaine02}, the mass of the host
bulge \citep{Magorrian98, Haring04, Graham04, Graham12}, and the luminosity of the host
bulge \citep{Kormendy95, Marconi03, Graham13}.  Furthermore, studies have found
that the activities of AGN and star formation are also related, both
peaking at similar redshifts and declining to the local downsized
Universe together \citep{Hopkins04, Silverman08, Aird10, Han12}.  All
this evidence suggests that SMBHs and their host galaxies share mutual
evolutionary histories \citep{Croton06, Bower06, DiMatteo05,
DiMatteo08, Hopkins05, Hopkins06, Hopkins08}.

While the observational evidence is still not conclusive, theoretical
studies have suggested two mechanisms responsible for the trigger of
starbursts and the ignition of AGN activity.  In the early universe,
major-mergers provide abundant fuel for the brightest quasars as
having been observed \citep[e.g.,][]{Treister12}, leading
to their space density peaking at $z = $2-3.  In the mean time, a
dusty starburst is also triggered by the mergers and the star
formation rate peaks at the same redshift range \citep[e.g.,
][]{Bouwens09}.  In a later stage below $z \sim 1$, AGN cosmic
downsizing happens \citep[e.g., ][]{Heckman04} in which the space
density of low-luminosity AGNs peaks at these redshifts because
secular evolution and minor interactions become the main fueling
mechanisms.

Despite the fact that an evolutionary sequence is still unclear, both massive
starburst and AGN activity can co-exist and heat the luminous
infrared galaxies (LIRGs, $L_{IR} = 10^{11}-10^{12}L_\sun$) and ultra
luminous infrared galaxies (ULIRGs, $L_{IR} > 10^{12}L_\sun$)
\citep[e.g., ][]{Sanders1988, Magnelli11, Murphy11}.  As LIRGs and
ULIRGs evolve, the AGN, instead of the dusty starburst, begins to dominate
the IR emission and heats the dust to higher temperatures, eventually
destroying or excavating the dust and gas associated with the starburst,
not only quenching the star formation, but also halting the black hole
growth and limiting the SMBH mass \citep{DiMatteo05}.  When the AGN runs out
of fuel and stops being active, SMBHs are left in the center of
massive galaxies.

Post-starburst quasars (PSQ)
are objects which show simultaneously an AGN and a massive
luminous post-starburst stellar population, with composite spectra displaying broad emission
lines as well as the Balmer jumps and strong Balmer absorption lines
characteristic of type-A stars.  The prototype of PSQ is UN~J1025-0040
\citep{Brotherton99, Canalizo00, Brotherton02}.  Its strong stellar
component has an age of $\sim 400$ Myr, and a bolometric luminosity comparable to that of the quasar.
A younger UN~J1025-0044 would have a more luminous stellar population and would
likely be dust enshrouded, placing it in the ULIRG class.  This object was proposed as 
representative of ULIRGs making an evolutionary transition into a quasar phase.

The large quasar surveys like the Sloan Digital Sky Survey (SDSS) have
made statistical studies of PSQs possible.  A sample of luminous
PSQs, spectroscopically selected from SDSS, at $z \sim 0.3$, were
studied via HST/ACS F606W imaging (\citealt{Cales11}, hereafter
C11) and Keck/KPNO high signal-to-noise ratio (S/N) optical
spectroscopy (\citealt{Cales13}, hereafter C13).  These studies showed
that PSQs have a heterogeneous population of both early-type and spiral
hosts.  Although these two subsets have similar disturbance fractions
in their hosts, the starburst mass and age of their hosts and the AGN
SMBH mass and Eddington fractions are different.  They concluded that
the early-type PSQs likely result from major mergers and are evolved
(U)LIRGs, while spiral PSQs imply more complicated star-formation history
and are likely triggered by more common non-merger events (e.g.,
harassment, bars).  The two types of PSQs with different AGN and
post-starburst signatures suggests that at least two mechanisms are
responsible for their triggering, supporting the relevant theoretical
studies involving mutual BH-bulge growth \citep[e.g.,][]{Hopkins09,
Schawinski10}.  Only the PSQs with early-type
hosts, however, appear to represent transitioning objects in the classic 
evolutionary sequence involving major mergers.  They also note that the 
PSQs with spiral hosts show morphological features and emission-line 
ratios indicating they possess ongoing or recent star formation in addition
to a dominant post-starburst stellar population.

We continue the study of the PSQs from C11 and C13 here, in which we 
investigate the mid-infrared (MIR) spectroscopic properties of
16 PSQs using {\it{Spitzer}} IRS observations.  We compare their MIR spectral 
features with host morphologies, MIR colors, and other properties
in an effort to better understand PSQs and their relationships to 
quasars and starburst galaxies more generally.

The MIR continuum is emitted by warm dust
heated by high-energy photons from an AGN, star formation, or
some mix of the two.  
Since the MIR continuum emission may be a combination of
thermal radiation from dust with different temperatures, its shape
and slope can help us understand the energy source.  Below
20\,\micron\ the continuum can be represented with a power law
($\nu^{-2}$, \citealt{Mullaney11}), and it flattens at longer
wavelengths.

In addition to information contained in
the continuum, several MIR spectral features can
be used to diagnose the relative AGN/starburst contributions to the
total infrared luminosity.

First, polycyclic aromatic hydrocarbon (PAH) emission lines exhibited
in the MIR spectrum can indicate intense star formation,
because PAHs are excited by UV and/or optical photons mostly from
young stars instead of from AGNs \citep[e.g., ][]{Peeters04,
Brandl06}.  Based on early {\it{Spitzer}} observations, several studies
\citep{Schweitzer06, Netzer07, Shi07, Hao07} have reported the
detection of PAH emission features in quasars selected with different
properties, including optically blue PG quasars, 2MASS red
quasars, and 3CR radio quasars.  This suggests that star formation
occurs widely in quasars and may also be present in PSQs.

Second, The 9.7\,\micron\ and 18\,\micron\ silicate absorption or emission
features are present in the mid-IR spectrum of quasars \citep[e.g.,
][]{Hao07}.  The silicate emission is suggested to arise mainly from
the inner edge of the dusty AGN torus according to AGN unified
schemes \citep[e.g.,][]{Antonucci93,Urry95}, while the absorption features may be produced by a dust screen
surrounding a hot emission region.

Our sample of PSQs is described in Section \ref{sec:sample}.  Details
of the observations and data reduction are explained in Section
\ref{sec:obs-data}.  
We describe the measurements of the PSQs spectral features in Section
\ref{sec:meas}. The mid-infrared composite spectrum of PSQs are given
in Section \ref{sec:comp}.  Results and discussions are given in Section
\ref{sec:result} .  Finally we summarize our results in Section
\ref{sec:summary}.  We adopt cosmological parameters
$\emph{H}_{0}=70\,\kms$~Mpc$^{-1}$, 
$\Omega_{m}=0.3$, and $\Omega_{\Lambda}=0.7$
throughout this paper.

\section{SAMPLE\label{sec:sample}}

Our sample was drawn from the HST snapshot program described by C11.
There were 80 objects in the program and we eventually obtained HST images
for 29 objects.
Before the HST snapshot program had been completed, we proposed and
obtained {\it{Spitzer}} IRS observations of the 16 objects that
currently had HST imaging.
These 16 objects are studied in this paper and
they are also included in a subsample of 38 PSQs studied with  high
S/N optical spectroscopy for separating the AGN and
stellar components (C13).

We summarize the selection criteria here, but see also C11 for more details.
All PSQs were spectroscopically selected from the Sloan Digital
Sky Survey data release 3 \citep[SDSS DR3, ][]{Abazajian05}, and
all objects have broad optical emission lines with FWHM $> 1000 \kms$
and significant detections of Balmer absorption lines and Balmer jumps
indicative of intermediate-age stellar populations (100s of Megayears).
Additionally all objects have $0.25 < z <0.45$ and $M_{r} \sim$\ -22.9.

Given our selection method
for the 80 HST targets and the random selection of the SNAP
observations, the 16 targets are likely representative for the class within
the redshift range.  Despite this small sample size, the luminosity of the
sample spans 1.5 dex, relatively significant for statistical studies.

\section{OBSERVATION AND DATA REDUCTION\label{sec:obs-data}}

Mid-infrared spectra of 16 PSQs were obtained with the {\it{Spitzer}}
Infrared Spectrograph \citep[IRS, ][]{Houck04} as part of our GO
program (Program ID 30075).  All 16 of the objects were observed in
staring mode with the Short Low first-order module (SL1 \& SL2), which
covers 5-14\,\micron, and with the Long Low second-order module (LL1 \&
LL2), which covers 14-40\,\micron.  The Ramp Duration and Number of
Cycles of the observations are listed in Table~\ref{tb:obs}.  The
{\it{Spitzer}} IRS has a slit width of 3\arcsec.6 for SL and 10\arcsec.7 for
LL, corresponding to 16 kpc and 47 kpc at $z=0.3$, respectively.  The
angular size of the majority of our objects is consistent with the expected
PSF size, thus we treated them as point sources.

We started with the standard pipeline products of co-added,
non-subtracted post-BCD frames.  The bad pixels in the images were
removed with the interactive IDL procedure IRSCLEAN (version 2.1.1).
Subtraction of 2-dimensional images with different nod positions of 
the same spectral order was used for sky subtraction.    
One-dimensional spectra were extracted with {\it{Spitzer}}
IRS Custom Extraction software (SPICE; version 2.5).  We adopted the
optimal extraction method for point sources with default parameters.
During this extraction, pixels are given different weights based on
their positions and a spatial profile of a bright calibration star.
Standard IRS flux calibration is applied, which also corrects slit
losses for unresolved sources.  The uncertainty of flux calibration is
within $\sim$ 5\%.  For each object, we obtained 6 individual
1-dimensional spectra from each of the two nod positions.
Corresponding spectra of the two nod positions agree well with each
other in both flux level and spectral shape.  The ``bonus'' 3rd order
short spectra have very low S/N and were discarded.  We averaged all
other individual spectra to obtain the final spectrum of each source
over the entire wavelength range.

We specifically checked the region near 14\,\micron\ where different
IRS spectral orders are stitched.  We found no flux excess in the
wider-slit LL modules compared to the SL modules, confirming that
there is no substantial slit loss and our objects can be treated as
point sources.  We also compared our spectra with those produced from
the standard Post-BCD pipeline and found that the flux level matches
well and ours have higher S/N due to the optimal extraction.

\section{SPECTRAL MEASUREMENTS\label{sec:meas}}

Table~\ref{tb:mir} lists our measurements of the PAH line fluxes and
equivalent widths (EWs), silicate strength defined by \citet{Hao07},
integrated fluxes around 5.5\,\micron\ and 15\,\micron.  The PAH
features are measured by fitting a single Gaussian profile and a local
continuum between 5.5\,\micron\ to 6.9\,\micron\ for 6.2\,\micron\ PAH
and between 11.0\,\micron\ to 11.7\,\micron\ for 11.3\,\micron\ PAH.
When the S/N is too low to measure, we estimated a 3$\sigma$\ upper
limit by simulating a Gaussian emission profile with the typical PAH
FWHM of 0.2\,\micron\ \citep{Smith07} of the low-resolution observing
mode.  We cannot obtain a clean measurement for the PAH 7.7\,\micron\
feature because of the complicated line blending in that region and
low S/N of our spectra.

\begin{figure}
\plottwo{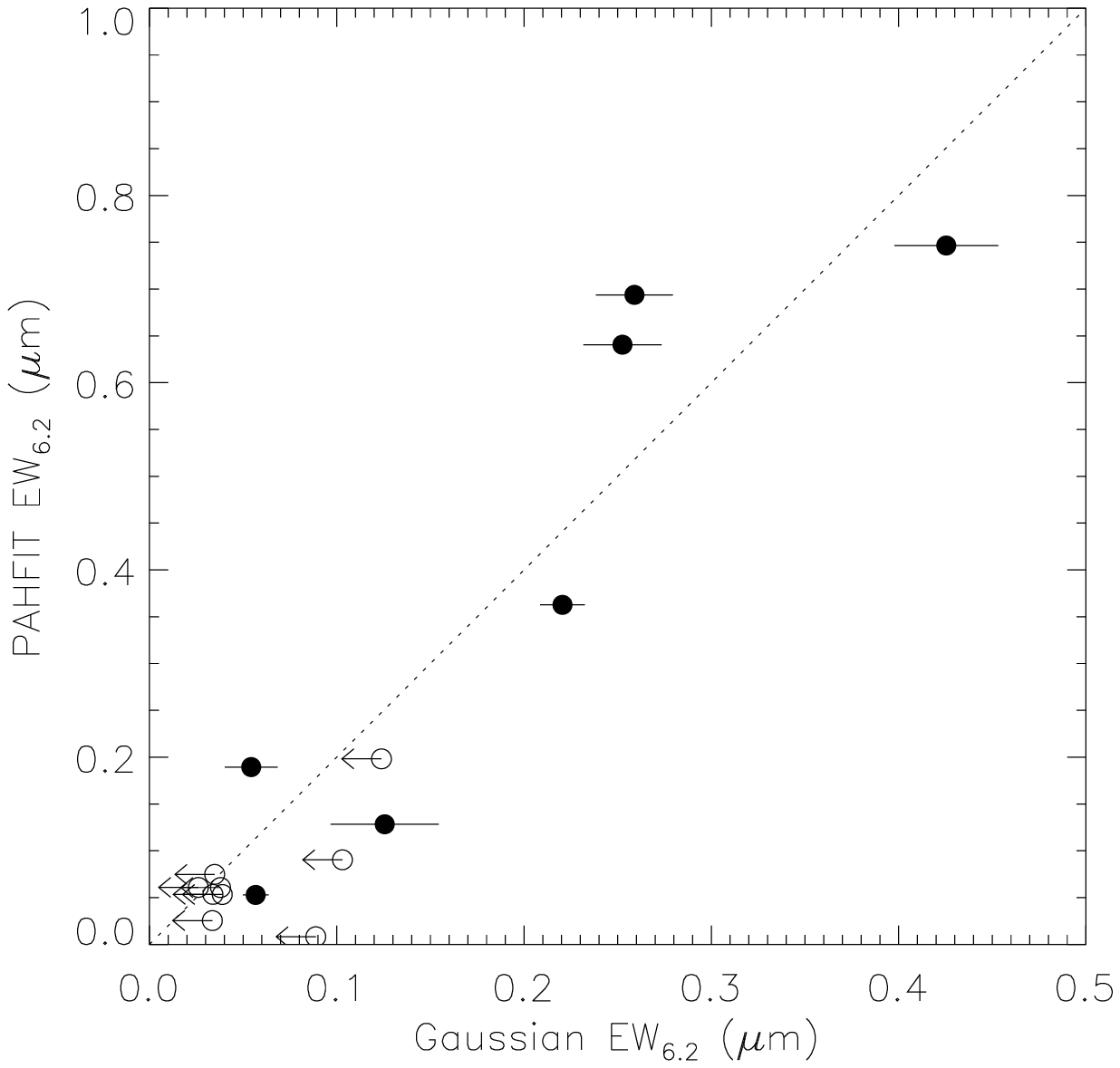}{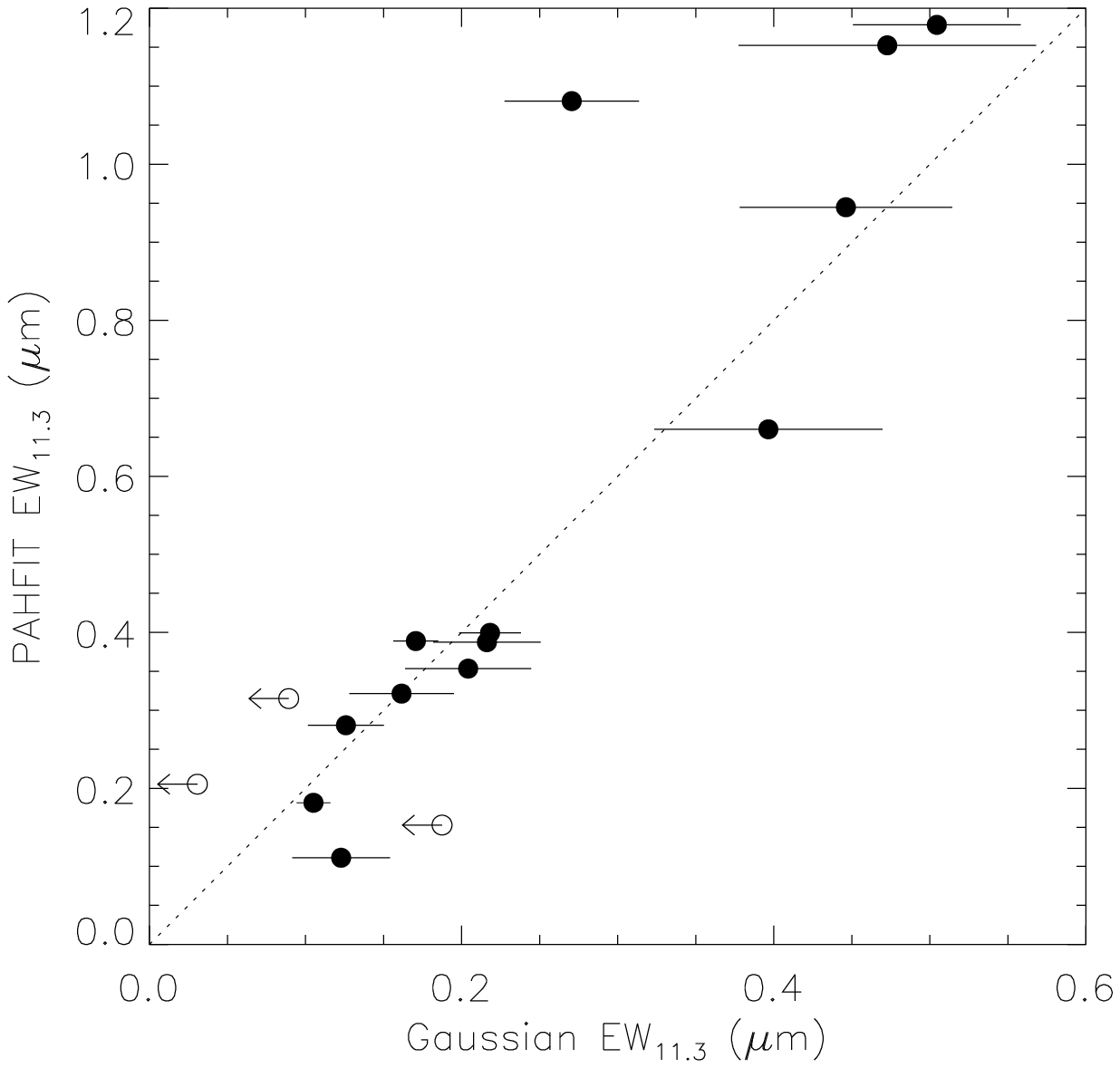}

\caption{Comparison of the PAH measurements using PAHFIT and 
Gaussian profile fit.
The dotted lines indicate a 2:1 relationship between the
two measurements.  The open symbols indicate upper limits from the
Gaussian profile fit (\S\ref{sec:meas}).
\label{fg:pahfit}}

\end{figure}

Since the S/N of our spectra is relatively low, there is probably no
sufficient information for detailed modeling such as PAHFIT
\citep{Smith07}, which can decomposes the spectra using physically
motivated models involving PAH emissions, gas temperatures, narrow
emission lines, dust absorptions, etc.  Our comparison shows that the
PAH EWs measured using PAHFIT are roughly twice as large as those
measured using Gaussian profile fit (Fig.~\ref{fg:pahfit}), because
part of the Gaussian fit continuum is accounted for by the broad wings
of PAH emission models in PAHFIT.  However, the PAH EWs measured using
the two methods are strongly correlated for our sample and therefore
the choice of adopting either measurement in the analyses does not
affect the conclusions.

Table~\ref{tb:mir} includes some additional information of interest.
We obtained the Wide-Field Infrared Survey Explorer
\citep[WISE,][]{Wright10}
profile-fit magnitudes and
uncertainties of our sources from the WISE All-Sky Release Point
Source Catalog \citep{Cutri12}.  We tabulate the WISE $W1-W2$ (i.e., [3.4\,\micron] $-$
[4.6\,\micron])\ MIR color here in 
column 9.  Column 10 provides host
morphology classification from \citet{Cales11}.  

\section{RESULTS AND DISCUSSION\label{sec:result}}

\subsection{Composite Spectra\label{sec:comp}}

\begin{figure}
\epsscale{.60}
\plotone{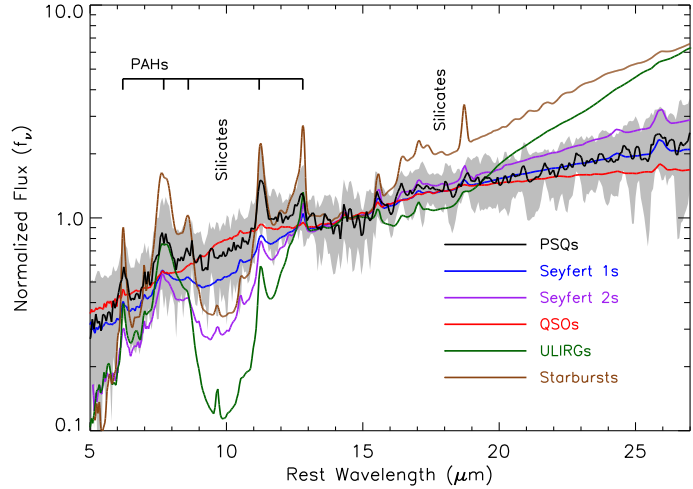}

\caption{Composite MIR low-resolution spectra of PSQs (black solid
line) in our sample, and QSOs (red), Seyfert 1s (blue), Seyfert 2s
(purple), ULIRGs (green) from \citet{Hao07}, and starbursts (brown) from
\citet{Brandl06}. They are normalized at 14.5\,\micron.
Shaded area indicates the 1-$\sigma$ dispersion of individual spectra
of our sample.  \label{fg:comps-all}}

\end{figure}

Because we are interested in PSQ ensemble as well as individual
characteristics, we construct composite spectra. 
First, we created a composite 
of the total sample of 16 PSQs by normalizing
their spectra at rest-frame 14.5\,\micron\ and applying a median
combine.  In Fig.~\ref{fg:comps-all}, we compare our PSQ composite
spectrum with those of Seyferts, PG QSOs, and ULIRGs \citep{Hao07} as
well as starburst \citep{Brandl06}.  It is clear that the PSQs lie
between ULIRGs and QSOs, especially between 15\,\micron\ to 30\,\micron\
where it is closer to Seyfert~1.  While type I QSOs show emission in
the 9.7\,\micron\ silicate feature, meanwhile ULIRGs and type II AGNs show
absorption, PSQs do not seem to show either emission or absorption.
In addition, the PAH features in PSQs are stronger than those in type
I AGNs, but weaker than those in starbursts.  These comparisons are consistent with
the idea that PSQs are hybrids of AGN and starbursts as also seen in optical
spectra.

\begin{figure}
\epsscale{.60}
\plotone{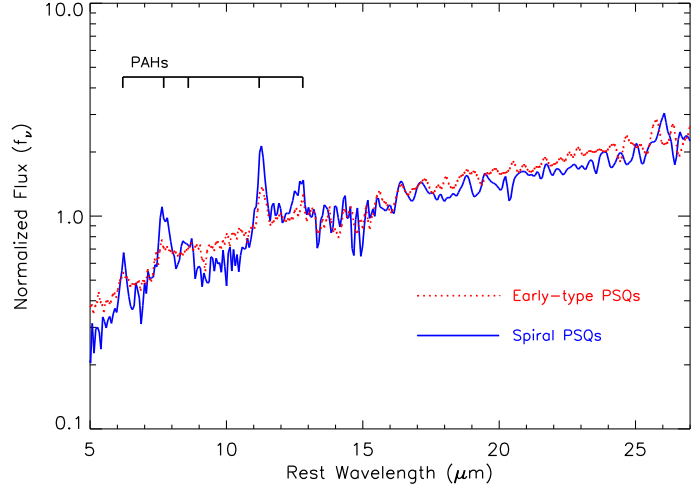}

\caption{Composite spectra of PSQs with 
early-type hosts (red dotted line) and spiral hosts (blue solid line), 
normalized at 14.5\,\micron.  \label{fg:comps-type}}

\end{figure}

We have also constructed composite spectra based on the host 
morphologies.  For the 16 PSQs in our sample, there are
8 with early-type hosts, 7 with spiral hosts, and one with
indeterminate
classification which is not included in constructing composite
spectra.
Fig.~\ref{fg:comps-type} shows the two composite spectra.
The continuum shapes show a
remarkable similarity, indicating a very small variation of the MIR
SEDs of PSQs residing in different types of host galaxies.  A slight
difference is that the PAH emission features are stronger in PSQs with
spiral hosts, suggesting more star formation activity.

\subsection{PSQs in Diagnostic Diagrams\label{sec:diagnostic}}

Several MIR diagnostic diagrams have been developed to study the
nature of the infrared emission \citep[e.g., ][]{Genzel98, Lutz98,
Rigopoulou99, Laurent00, Peeters04, Spoon07}.  The strength of the PAH
features can be used to distinguish the dominant energy source, AGN or
stellar; and the silicate absorption or emission can be used to
diagnose obscured or unobscured sources.

\begin{figure}


\epsscale{.50}
\plotone{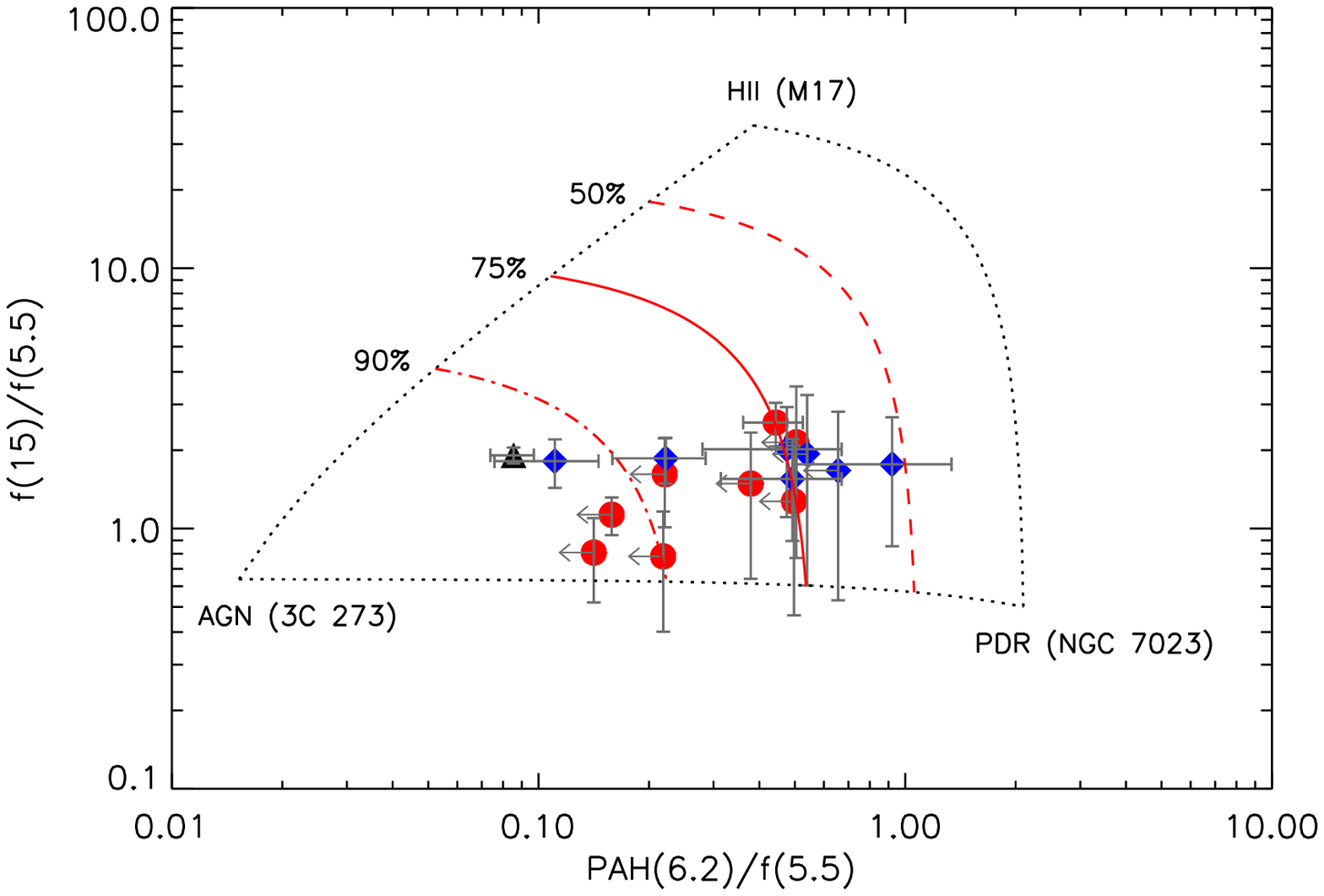}

\caption{Mid-infrared diagnostic diagram as in \citet{Armus07}, which
is evolved from that of \citet{Laurent00},  involving
the integrated continuum fluxes $f(15)$ (from 14-16\,\micron) and
$f(5.5)$ (from 5.3-5.8\,\micron), and the 6.2\,\micron\ PAH flux.  Three
extreme cases are labeled at the three vertices.  The red lines
indicate the fractional AGN contribution.  Red circles represents our
PSQs in early-type host galaxies, blue diamonds the spiral hosts, and
grey triangle the indeterminate host.  It implies that PSQs are
intermediaries between AGN and starbursts.
\label{fg:diagnos-l2000}} 

\end{figure}

Figure~\ref{fg:diagnos-l2000} shows our objects in the diagnostic
diagram of \citet{Armus07} where the flux ratio at 15\,\micron\ and
5.5\,\micron\ ($f_{15}/f_{5.5}$) and the ratio of the 6.2\,\micron\
PAH flux to $f_{5.5}$ are used to distinguish the
dominant type of emission region: AGN, starburst (HII) or
photodissociated region (PDR).  All our PSQs fall in the middle of the
diagram, indicating they are intermediate between AGN and starburst,
but with all having more than 50\% contribution from AGNs.

The ``fork diagram'' of \citet{Spoon07}, simply involving the EW of
PAH 6.2\,\micron\ (EW$_{6.2}$) and the strength of the 9.7\,\micron\
silicate feature, provides better diagnostics.  We reproduced the
diagram in Figure~\ref{fg:diagnos-spoon07} using data from
\citet{Spoon07} and added the PSQs.  It is clear that hyperluminous
infrared galaxies (HyLIRGs) and ULIRGs show more silicate absorption
because of their dust rich environment, while unobscured QSOs suppress
PAH emission.
Again, PSQs have the MIR properties between pure AGNs and pure
starbursts, consistent with the above and the evidence from the
composite spectra (\S~\ref{sec:comp}).  
Moreover, the silicate strength of PSQs is very weak and uncertain,
showing no obvious absorption (or emission) and suggesting that PSQs
are probably not obscured, or at least not heavily obscured.
Both the host galaxy and the circumnuclear dust torus can be
responsible for the absorption of the silicate features, and thus the
obscuration of the PSQs.  \citet{Goulding12} have studied a sample of
Compton-thick AGNs and suggested that the silicate absorption seems to
originate from the host galaxy.  However, with current data, we cannot
address whether the lack of obscuration of our objects is due to the
host galaxy or the torus.

When studying 70 luminous infrared galaxies ($z \sim 0.5-3$),
\citet{Hernan09} also compiled a reference sample of 137 low-redshift
sources including quasars, type 1 and 2 Seyfert galaxies, ULIRGs and
starbursts.  They compared different MIR diagnostic diagrams using the
combined sample of more than 200 sources and found that EW$_{6.2}=0.2$
separates most starburst-dominated and AGN-dominated sources.  This
criterion would place most PSQs as AGN-dominated.  While this
definition is somewhat arbitrary, we do notice when using this
definition that PSQs in early-type hosts are more AGN-dominated and
those in spiral hosts seem to be closer to starbursts.

Based on all the evidence from the PSQ composite spectrum and
diagnostic diagrams, we can infer that the dust in PSQs is heated by
both AGN and stellar process.
The PAH emission strengths of PSQs are higher than those of PG QSOs,
but similar to those of ULIRGs, as seen in
Figure~\ref{fg:diagnos-spoon07}.  This may imply that PSQs are more
similar to ULIRGs than to QSOs, except that PSQs do not show obvious
silicate absorptions as ULIRGs, indicating that PSQs are unobscured
AGNs.  But on the other hand, the fact that the PAH molecules in PSQs
can survive the high energy photons from the central AGN suggests that
they must be shielded by a large amount of gas and dust from radiation
\citep[see][]{Schweitzer06}.  It is therefore reasonable to suggest
that PSQs have less dust absorption than ULIRGs, but perhaps more than
QSOs.  The PSQ properties revealed in the mid-IR agrees with those
from the optical, especially supporting the hypothesis in an
evolutionary scenario that the PSQs are in a transitional phase from
ULIRGs to classical QSOs as they emerge from their dust cocoons.

%

\subsection{Mid-IR Properties of PSQs and Host Galaxy Type}

PSQs in early-type and spiral hosts show different PAH strengths, as
seen in \S\ref{sec:comp} and \S\ref{sec:diagnostic}.  
We explore this difference more quantitatively below mainly with the
PAH 11.3\,\micron\ emission feature which has higher S/N for our
objects.

\begin{figure}

\epsscale{.60}
\plotone{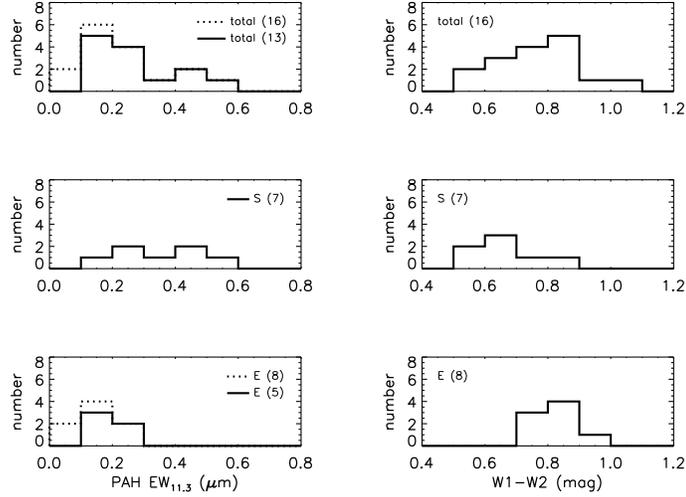}

\caption{Distributions of the 11.3\,\micron\ PAH EWs (left)
and the WISE $W1-W2$ mid-infrared colors (right) for the entire sample 
(top), 7 spiral hosts PSQs (middle), and 8 early-type
host PSQs (bottom).  For PAH EW$_{11.3}$, the dotted line and 
solid line indicate including and excluding the objects 
with upper limits, respectively.
\label{fg:hist}}

\end{figure}

Fig.~\ref{fg:hist} shows the distributions of PAH EW$_{11.3}$ for the
entire PSQ sample, the spiral-host PSQs, and the early-type host PSQs.  
The spiral-host PSQs have statistically stronger PAH EW$_{11.3}$ (mean
EW$_{11.3}= 0.35\pm0.13\micron$) than those with early-type host (mean
EW$_{11.3}<0.14\,\micron$, including 3 upper limits). 
This can also be seen clearly in Fig.~\ref{fg:wise-pah},
implying that spiral-host PSQs have more ongoing or recent star formation.
%

%

\citet{Stern12} identified a sample of 130 AGNs in the COSMOS field
\citep{Scoville07}
using a simple WISE mid-infrared color and suggested using WISE color
$W1-W2 \ge 0.8$ as a good criterion to select both unobscured and
obscured AGNs.  This color selection criterion is independent of
optical colors, radio and X-ray properties of their selected AGNs
whose redshifts span between $0-3$.
For our objects with similar redshifts, we can also use $W1-W2$ as a
tracer of AGN dominance.  Fig.~\ref{fg:hist} shows the distributions
of $W1-W2$.  The difference between the two host types are obvious
with PSQs in early-type hosts showing greater AGN contribution,
confirming the results from the optical work (C11, C13) that AGNs of
PSQs hosted by early-type galaxies have on average greater AGN
luminosity than those hosted by spiral galaxies.  
Fig.~\ref{fg:wise-pah} also shows distinct regions for the two host
types, as well as a strong anticorrelation between PAH EW$_{11.3}$ and
$W1-W2$.  At least two factors may be responsible for this
anticorrelation.  First, continuum emission from warm dust rises with
increasing AGN activity and therefore dilutes the EW of PAH emission
even if the flux of the PAH emission remains the same.  Second, PAH
molecules can be destroyed by high energy photons \citep[e.g.,][]{Voit92}
with increasing AGN
activity as indicated by higher $W1-W2$.

Our results are consistent with many previous studies that have
compared MIR spectral properties with optical diagnostics and
specifically investigated the connections between the PAH emissions,
star formation, and AGN activities, using different samples in the
local ($z < 0.2$) universe
\citep[e.g.,][]{ODowd09,Treyer10,LaMassa12}.
They found that AGN dominated sources have weaker PAH EWs and if this
is caused by the ionization field hardness, it is only true for the
AGN population and cannot be generalized for the star-forming
population.  While our small sample, with higher luminosity and higher
redshift (compared to $z<0.1$ for most Seyfert 2s in
\citet{LaMassa12}), also shows weaker PAH emission with increasing AGN
activity, the data quality of our sample does not allow us to address
the issue of the hardness of radiation fields.


We have also looked for correlations between MIR properties and derived
parameters of AGN (black hole mass and Eddington accretion rate) or
starburst (age and mass) that were obtained from the optical spectral
decompositions in C13, and found no significant relationships of interest.  No correlation
was seen between optical AGN and starburst parameters either (C13), which
is perhaps surprising.  We note that there may be several reasons for
this.
1. Our sample is small and the uncertainty of the measured parameters
due to low S/N may have hidden the expected connections.
2. The derived parameters of AGN and starburst from the optical
spectral decomposition may have intrinsic uncertainty due to possible
degeneracy issues in the spectral modeling.
3. There does not exist any correlation between these parameters which
leads to the inevitable thought that the dust structure of PSQs, the
interaction between black hole, starburst, and dust may be more
complicated than we thought.

\section{SUMMARY\label{sec:summary}}

We have studied the MIR spectral properties of 16 PSQs based on
observations with {\it{Spitzer}} IRS, trying to better understand the nature
of the PSQs.  All of our analyses show consistent results.

1. MIR spectra show supporting evidence that PSQs are hybrid AGN and
starburst systems as seen in the optical spectra.

2. Diagnostic diagrams as well as all detailed MIR features in the PSQ
composite spectrum, including continuum slope, silicate features, and
PAH emission features (6.2\,\micron\ and 11.3\,\micron), seem to place PSQs
in between ULIRGs and normal QSOs, supporting the hypothesis that some
PSQs are in the transition phase from ULIRGs evolving to QSOs (e.g. those
in early type hosts) or simply hybrid objects triggered by events less
dramatic than major mergers (e.g., those in spiral hosts).

3. PSQs in early-type hosts have greater AGN activity,
consistent with the results from optical studies (C11).

4. PSQs in spiral hosts have stronger PAH emission, indicating that
they have more ongoing star formation.

\begin{figure}

\epsscale{.60}
\plotone{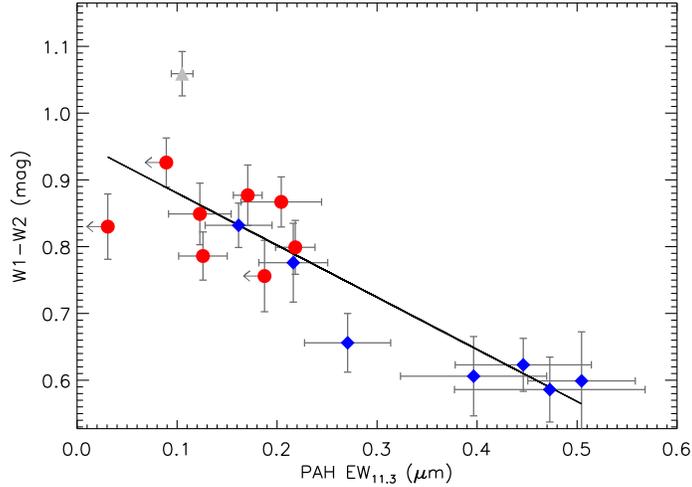}

\caption{Anticorrelation between WISE mid-infrared color $W1-W2$ and
11.3\,\micron\ PAH EWs for early-type host PSQs (red circles),
spiral host PSQs (blue diamonds), and the indeterminate host PSQ (grey
triangle).  The correlation coefficient is
-0.851 with a statistic significant level of $>99.997\%$.  The solid line
indicates the  best fit.  It is also clear that spiral-host PSQs have
in average larger PAH EWs than early-type PSQs. \label{fg:wise-pah}}


\end{figure}

5. PAH emission decreases with increasing AGN dominance in PSQs
(Fig.~\ref{fg:wise-pah}), suggesting that the PAH emission is either
diluted by AGN continuum or the PAH molecules are destroyed by high
energy photos from AGN.


These luminous $z \sim 0.3$ PSQs are interesting, extreme hybrid or
transitory objects.  Both optical and infrared diagnostics provide a
consistent picture in which we see two populations distinguished by
their host morphology, one likely representative of major-merger
triggers thought to be common in the so-called ``quasar epoch'' of $2
< z < 3$, and the other characteristic of the more local ``downsized
universe'' in which lower mass and lower luminosity systems are active
and triggered by less dramatic events.  PSQs seem to be useful as
objects capable of diagnosing the two paths toward AGN activity
associated with bursts of star formation.

\acknowledgments

\begin{figure}


\epsscale{.60}
\plotone{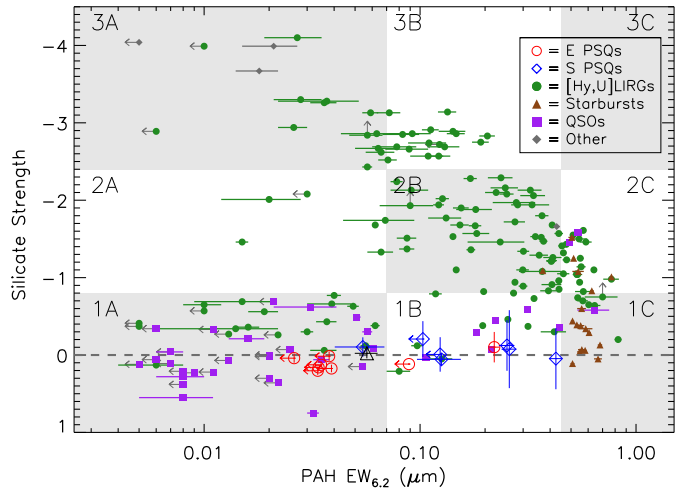}

\caption{Simplified ``Fork diagnostic diagram'' \citep{Spoon07} with
our PSQs added, showing  the 9.7\,\micron\ silicate strength vs. the
equivalent width of the 6.2\,\micron\ PAH emission feature.
Red open circles are PSQs in early-type hosts and blue open diamonds
the PSQs in spiral hosts.  The PSQ in indeterminate host is indicated with
an open triangle.
Other galaxy types include ULIRGs and HyLIRGs (filled green circles),
starburst galaxies (brown triangles), Seyfert galaxies and QSOs
(filled purple
squares), and other infrared galaxies (filled grey diamonds).
A simple way to interpret the diagram is that starbursting increases
from left 
to right (strong PAH emission) and dust obscuration rises from bottom
to top (strong silicate absorption).
\label{fg:diagnos-spoon07}}

\end{figure}

We thank the anonymous referee for constructive suggestions to improve
the paper.
We thank Henrik Spoon for providing us with the data for the
diagnostic diagram (Fig~\ref{fg:diagnos-spoon07}).  We acknowledge
support from the National Natural Science Foundation of China through
grant 10633040 and support by Chinese 973 Program 2007CB815405 and
Tianjin Distinguished Professor Funds.  M.S.B acknowledges support
from NASA through the LTSA grant NNG05GE84G.
S. L. Cales was supported in part by NASA Headquarters under the NASA
Earth and Space Science Fellowship Program (Grant NNX08AX07H), in part
by National Science Foundation GK-12 Program (Project 0841298) and
also in part by ALMA-CONICYT program 31110020.

This work is based on observations made with the Spitzer Space
Telescope, which is operated by the Jet Propulsion Laboratory,
California Institute of Technology under a contract with NASA.  This
publication makes use of data products from the Wide-field Infrared
Survey Explorer, which is a joint project of the University of
California, Los Angeles, and the Jet Propulsion Laboratory/California
Institute of Technology, funded by NASA.

{\it Facilities:} \facility{{\it{Spitzer}} (IRS) },\facility{WISE}.


\clearpage

\clearpage

%
%
%
%
%


\begin{deluxetable}{rlccrrr}
\tabletypesize{\scriptsize}
\tablecolumns{12}
\tablewidth{0pc}
\tablecaption{PSQ Sample and IRS Observing Parameters
\label{tb:obs}}
\tablehead{
\colhead{} &
\colhead{} &
\colhead{} &
\multicolumn{4}{c}{Ramp Duration (in sec) $\times$ Cycles} \\
\cline{4-7}
\colhead{ID} &
\colhead{Object Name} &
\colhead{$z$\tablenotemark{a}}&
\colhead{SL2} &
\colhead{SL1} &
\colhead{LL2} &
\colhead{LL1} \\
\colhead{(1)} &
\colhead{(2)}&
\colhead{(3)}&
\colhead{(4)}&
\colhead{(5)} &
\colhead{(6)} &
\colhead{(7)}
}
\startdata
1 & SDSS J020258.94$-$002807.5 & 0.339 & 60$ \times$4 & 60$ \times$2 &
120$ \times$6 & 120$ \times$6 \\
2 & SDSS J021447.00$-$003250.6 & 0.349 & 60$ \times$5 & 60$ \times$2 &
120$ \times$8 & 120$ \times$8 \\
3 & SDSS J023700.30$-$010130.5 & 0.344 & 60$ \times$4 & 60$ \times$2 &
120$ \times$5 & 120$ \times$5 \\
4 & SDSS J074621.06+335040.8 & 0.284 & 60$ \times$2 & 14$ \times$4 &
120$ \times$2 & 120$ \times$3 \\
5 & SDSS J075045.00+212546.3 & 0.408 & 60$ \times$2 & 14$ \times$3 &
120$ \times$2 & 120$ \times$2 \\
6 & SDSS J075521.30+295039.2 & 0.334 & 60$ \times$5 & 60$ \times$2 &
120$ \times$7 & 120$ \times$8 \\
7 & SDSS J075549.56+321704.1 & 0.420 & 60$ \times$6 & 60$ \times$2 &
120$ \times$9 & 120$ \times$9 \\
8 & SDSS J081018.67+250921.2 & 0.263 & 14$ \times$2 & 6$ \times$2 &
14$ \times$3 & 14$ \times$3 \\
9 & SDSS J105816.81+102414.5 & 0.275 & 60$ \times$3 & 14$ \times$6 &
120$ \times$5 & 120$ \times$5 \\
10 & SDSS J115355.58+582442.3 & 0.319 & 60$ \times$5 & 60$ \times$2 &
120$ \times$4 & 120$ \times$5 \\
11 & SDSS J124833.52+563507.4 & 0.266 & 14$ \times$2 & 6$ \times$2 &
14$ \times$2 & 14$ \times$2 \\
12 & SDSS J145640.99+524727.2 & 0.277 & 60$ \times$2 & 14$ \times$2 &
14$ \times$5 & 14$ \times$6 \\
13 & SDSS J154534.55+573625.1 & 0.268 & 60$ \times$2 & 14$ \times$2 &
14$ \times$6 & 30$ \times$4 \\
14 & SDSS J170046.95+622056.4 & 0.276 & 60$ \times$3 & 14$ \times$3 &
30$ \times$5 & 30$ \times$7 \\
15 & SDSS J212843.42+002435.6 & 0.346 & 60$ \times$4 & 60$ \times$2 &
120$ \times$6 & 120$ \times$6 \\
16 & SDSS J231055.50$-$090107.6 & 0.364 & 60$ \times$4 & 60$ \times$2
& 120$ \times$5 & 120$ \times$5 \\
\enddata
\tablenotetext{a}{Redshift from SDSS DR7}

\end{deluxetable}

\clearpage

\begin{deluxetable}{rrrcrrccccc}
\tabletypesize{\scriptsize}
\tablecolumns{11}
\tablewidth{0pc}
\tablecaption{Measurements of the MIR Features \label{tb:mir}}
\tablehead{
\colhead{}  &
\multicolumn{2}{c}{PAH 6.2\,\micron} &
\colhead{} &
\multicolumn{2}{c}{PAH 11.3\,\micron} &
\colhead{} & \colhead{} & \colhead{} & \colhead{} &
\colhead{Morphology } \\
\cline{2-3} \cline{5-6}
\colhead{ID} &
\colhead{Flux\tablenotemark{a}}&
\colhead{EW\tablenotemark{b}}&
\colhead{} &
\colhead{Flux\tablenotemark{a}}&
\colhead{EW\tablenotemark{b}}&
\colhead{Silicate 9.7\,\micron \tablenotemark{c}} &
\colhead{f$_{5.5}$\tablenotemark{d}} &
\colhead{f$_{15}$\tablenotemark{e}} &
\colhead{$W1-W2$\tablenotemark{f}}&
\colhead{Classification} \\
\colhead{(1)} & \colhead{(2)}& \colhead{(3)}&
\colhead{} &
\colhead{(4)}& \colhead{(5)} & \colhead{(6)} &
\colhead{(7)} & \colhead{(8)} & \colhead{(9)} &
\colhead{(10)}
}
\startdata
1       & 0.137 & 0.057 &       &
0.207 & 0.105 & $-$0.022      & 1.600 & 3.070 & 1.059 & Indeterminate
\\
2       & 0.228 & 0.221 &       &
0.187 & 0.218 & $-$0.102      & 0.514 & 1.310 & 0.799 & Elliptical
\\
3       & 0.153 & 0.259 &       &
0.185 & 0.473 & $-$0.076      & 0.311 & 0.482 & 0.586 & Spiral
\\
4       & $<$ 0.103     & $<$ 0.038
&       & 0.108 & 0.171 &\ \  0.015 & 0.466 & 0.753 & 0.877 & Elliptical
\\
5       & $<$ 0.139     & $<$ 0.035
&       & 0.150 & 0.204 &\ \  0.126 & 0.879 & 0.994 & 0.867 & Elliptical
\\
6       & 0.194 & 0.425 &       &
0.158 & 0.504 &\ \  0.047 & 0.210 & 0.372 & 0.599 & Spiral        \\
7       & 0.062 & 0.126 &       &
0.070 & 0.216 &\ \  0.055 & 0.278 & 0.517 & 0.776 & Spiral        \\
8       & $<$ 0.376     & $<$ 0.089
&       & $<$ 0.170     & $<$ 0.089     &\ \  0.117 & 0.992 & 1.480 &
0.926 & Elliptical    \\
9       & $<$ 0.118     & $<$ 0.034
&       & $<$ 0.066     & $<$ 0.187     &\ \  0.203 & 0.237 & 0.302 &
0.756 & Elliptical    \\
10       & $<$ 0.065     & $<$ 0.034
&       & $<$ 0.037     & $<$ 0.031     &\ \  0.160 & 0.295 & 0.231 &
0.830 & Elliptical    \\
11      & $<$ 0.405     & $<$ 0.039
&       & 0.199 & 0.126 &\ \  0.174 & 0.803 & 1.720 & 0.786 & Elliptical
\\
12      & $<$ 0.158     & $<$ 0.124
&       & 0.124 & 0.271 & $-$0.008      & 0.241 & 0.402 & 0.656 &
Spiral        \\
13      & 0.079 & 0.054 &       &
0.108 & 0.162 & $-$0.104      & 0.710 & 1.290 & 0.832 & Spiral
\\
14      & $<$ 0.094     & $<$ 0.103
&       & 0.109 & 0.446 & $-$0.207      & 0.173 & 0.335 & 0.623 &
Spiral        \\
15      & $<$ 0.062     & $<$ 0.026
&       & 0.033 & 0.123 &\ \  0.040 & 0.438 & 0.354 & 0.849 & Elliptical
\\
16      & 0.136 & 0.253 &       &
0.162 & 0.397 & $-$0.123      & 0.287 & 0.579 & 0.606 & Spiral
\\
\enddata

\tablenotetext{a}{Observed-frame flux in the units of $10^{-20}$
\fluxw.  The 3$\sigma$ upper limits are
provided for nondetections.}

\tablenotetext{b}{Rest-frame equivalent width 
in the units of \micron.}

\tablenotetext{c}{Silicate strength at 9.7\,\micron\ defined by
\citet{Hao07}.}

\tablenotetext{d}{Integrated continuum flux from rest-frame
5.3-5.8\,\micron\ in the units of $10^{-20}$ \fluxw.}

\tablenotetext{e}{Integrated continuum flux from rest-frame
14-16\,\micron\ in the units of $10^{-20}$ \fluxw.}

\tablenotetext{f}{WISE mid-infrared colors $W1-W2$ (i.e.,
$[3.4\,\micron]-[4.6\,\micron]$) in the units of magnitude.}

\end{deluxetable}

\clearpage

\end{document}